\if00
\documentclass[conference,a4paper]{IEEEtran}
\else
\documentclass[10pt,draftclsnofoot,onecolumn]{IEEEtran}
\fi

\usepackage[table,dvipsnames]{xcolor}
\usepackage{cite}
\usepackage{soul}
\usepackage{tikz}
\usepackage[normalem]{ulem}
\usepackage{array}
\usepackage{float}
\usepackage{amsthm}
\usepackage{amsmath}
\usepackage{amssymb}
\usepackage{optidef}
\usepackage{enumitem}
\usepackage{graphicx}
\usepackage{pgfplots}
\usepackage{thmtools}
\usepackage{clipboard}
\usepackage{microtype}
\usepackage{tikzscale}
\usepackage{algorithm}
\usepackage{mathtools}
\usepackage{subcaption}
\usepackage{algorithmic}
\usepackage{thm-restate}

\usepackage[T1]{fontenc}
\usepackage[font=small,labelfont=bf]{caption}

\pgfplotsset{compat=newest}
\usetikzlibrary{arrows}
\usetikzlibrary{patterns}
\usetikzlibrary{plotmarks}
\usetikzlibrary{arrows.meta}
\usetikzlibrary{positioning}
\usepgfplotslibrary{patchplots}
\usetikzlibrary{external}
\allowdisplaybreaks

\theoremstyle{general} 
\theoremstyle{general} 
\theoremstyle{general} 
\theoremstyle{general} 
\theoremstyle{general} \newtheorem{p-corollary}{Corollary}[proposition]
\theoremstyle{general} 
\theoremstyle{general} 
\theoremstyle{remark}  \newtheorem{remark}{Remark}

\begin{document}
	\title{Successive Null-Space Precoder Design for Downlink MU-MIMO with Rate Splitting and Single-Stage SIC}
	\author{\IEEEauthorblockN{Aravindh Krishnamoorthy\rlap{\textsuperscript{\IEEEauthorrefmark{2}\IEEEauthorrefmark{1}}}\,\,\,  and Robert Schober\rlap{\textsuperscript{\IEEEauthorrefmark{2}}}\\
	\IEEEauthorblockA{\small \IEEEauthorrefmark{2}Friedrich-Alexander-Universit\"{a}t Erlangen-N\"{u}rnberg, Germany\\
	\IEEEauthorrefmark{1}Fraunhofer Institute for Integrated Circuits (IIS) Erlangen, Germany}}}
	\maketitle
	
	%%%%%%%%%%%%%%%%%%%%%%%%%%%%%%%%%%%%%%%%%%%%%%%%%%%%%%%%%%%%%%%%%%%%%%%%%%%%%%%%
	\begin{abstract}
		In this paper, we consider the precoder design for an under-loaded or critically loaded downlink multi-user multiple-input multiple-output (MU-MIMO) communication system. We propose novel precoding and decoding schemes which enhance system performance based on rate splitting at the transmitter and single-stage successive interference cancellation at the receivers. The proposed successive null-space (SNS) precoding scheme utilizes linear combinations of the null-space basis vectors of the successively augmented MIMO channel matrices of the users as precoding vectors to adjust the inter-user-interference experienced by the receivers. We formulate a non-convex weighted sum rate (WSR) optimization problem, and solve it via successive convex approximation to obtain a suboptimal solution for the precoding vectors and the associated power allocation. Our simulation results reveal that the proposed SNS precoders outperform block diagonalization based linear and rate splitting designs, and in many cases, have a relatively small gap to the maximum sum rate achieved by dirty paper coding.
	\end{abstract}
	%%%%%%%%%%%%%%%%%%%%%%%%%%%%%%%%%%%%%%%%%%%%%%%%%%%%%%%%%%%%%%%%%%%%%%%%%%%%%%%%
	
	%%%%%%%%%%%%%%%%%%%%%%%%%%%%%%%%%%%%%%%%%%%%%%%%%%%%%%%%%%%%%%%%%%%%%%%%%%%%%%%%
	\section{Introduction}
	%%%%%%%%%%%%%%%%%%%%%%%%%%%%%%%%%%%%%%%%%%%%%%%%%%%%%%%%%%%%%%%%%%%%%%%%%%%%%%%%
	With the ever-increasing demand for mobile data services, spectral efficiency and user fairness remain at the forefront of the requirements for beyond 5th generation (5G) communication systems \cite{Saad2020}, \cite{You2020}. Motivated by the recent improvements of the signal processing capabilities of user terminals, linear precoding and successive interference cancellation (SIC) based schemes for multiple-input multiple-output (MIMO) systems such as MIMO non-orthogonal multiple-access (NOMA) \cite{Ding2017,Makki2020} and MIMO rate-splitting multiple-access (RSMA) \cite{Mao2018,Zhou2020,Dizdar2020,Dizdar2020a} have been proposed in the context of beyond 5G downlink communication. MIMO-NOMA can achieve high spectral efficiency and user fairness \cite{Ding2017}. However, for $K$ users, MIMO-NOMA requires in total $K(K-1)/2$ SIC stages at the receivers, which incurs a high complexity if there are more than a few users.
	
	On the other hand, multiple stages of rate splitting (RS) at the transmitter combined with linear precoding can achieve a performance close to that of dirty paper coding (DPC) for MIMO channels \cite{Li2018}, \cite{Li2020}. However, these schemes also have a high overall complexity due to the multiple stages of encoding at the transmitter and virtual MIMO or SIC based decoding at the receivers \cite{Li2020}. On the contrary, the multiple-input single-output (MISO)- and MIMO-RSMA schemes proposed in \cite{Mao2018,Zhou2020,Dizdar2020,Dizdar2020a} split the users' messages into only two parts: a private part and a common part, thereby enabling reception with a single stage of SIC, and considerably reducing the encoding and decoding complexity. Therefore, these schemes provide a favorable trade-off between performance and computational complexity. Furthermore, RSMA-based schemes were also shown to be robust to imperfect channel state information (CSI) \cite{Joudeh2016}, \cite{Mao2020}.
	
	For under-loaded or critically loaded downlink MIMO systems, where the combined number of receive antennas is smaller than or equal to the number of transmit antennas, block diagonalization (BD) allows for inter-user-interference (IUI)-free communication \cite{Spencer2004}. However, the achievable rate for BD has a large gap to DPC. Thus, RSMA-based extensions to BD were studied in \cite{Flores2019}, \cite{Flores2020}, and improvements in performance and resilience to imperfect CSI were demonstrated. However, the schemes in \cite{Flores2019}, \cite{Flores2020} utilized scalar common messages, fixed BD or regularized BD precoder matrices, and performed power allocation based on exhaustive search. 

	In this paper, we present a novel successive null-space (SNS) linear precoder design as a generalization of BD, and utilize an optimization based power allocation framework to maximize the weighted sum rate (WSR). The proposed SNS precoding scheme utilizes linear combinations of the null-space basis vectors of the successively augmented MIMO channel matrices of the users as precoding vectors to adjust the IUI experienced by the receivers and enhance the WSR. The main contributions of this paper are as follows.
	\begin{itemize}
		\item We present novel SNS precoding and decoding schemes for RS-based downlink multi-user (MU)-MIMO communication which require only a single stage of SIC at the receivers.
		\item We formulate an optimization problem for maximization of the WSR, and solve it via successive convex approximation (SCA) \cite{Razaviyayn2014} to obtain a suboptimal solution for the precoding vectors and the associated power allocation. The obtained solution provides a feasible lower bound on the performance of the proposed precoding and decoding schemes.
		\item We compare the obtained lower bound with the performance of DPC, MIMO-RSMA \cite{Flores2019}, and BD. Our simulation results show that the proposed design outperforms MIMO-RSMA \cite{Flores2019} and BD, and has a relatively small gap to DPC, especially at low to medium signal to noise ratios (SNRs).
	\end{itemize} 

	The remainder of this paper is organized as follows. In Section \ref{sec:sm}, we present the system model. The proposed SNS precoding and decoding schemes are provided in Section \ref{sec:pd}. WSR optimization is studied in Section \ref{sec:mwsr}. Simulation results are presented in Section \ref{sec:sim}, and the paper is concluded in Section \ref{sec:con}.
		
	\emph{Notation:} Boldface capital letters $\boldsymbol{X}$ and boldface lower case letters $\boldsymbol{x}$ denote matrices and vectors, respectively. $\boldsymbol{X}^\mathrm{T}$, $\boldsymbol{X}^\mathrm{H}$, $\mathrm{tr}\mkern-\thinmuskip\left(\boldsymbol{X}\right)$, and $\mathrm{det}\mkern-\thinmuskip\left(\boldsymbol{X}\right)$ denote the transpose, Hermitian transpose, trace, and determinant of matrix $\boldsymbol{X}$, respectively. $\mathbb{C}^{m\times n}$ and $\mathbb{R}^{m\times n}$ denote the sets of all $m\times n$ matrices with complex-valued and real-valued entries, respectively.  $\boldsymbol{I}_N$ denotes the $N\times N$ identity matrix, and $\boldsymbol{0}$ denotes the all zero matrix of appropriate dimension. The circularly symmetric complex Gaussian (CSCG) distribution with mean vector $\boldsymbol{\mu}$ and covariance matrix $\boldsymbol{\Sigma}$ is denoted by $\mathcal{CN}(\boldsymbol{\mu},\boldsymbol{\Sigma})$; $\sim$ stands for ``distributed as''. $\mathrm{E}\mkern-\thinmuskip\left[\cdot\right]$ denotes statistical expectation.
	
	%%%%%%%%%%%%%%%%%%%%%%%%%%%%%%%%%%%%%%%%%%%%%%%%%%%%%%%%%%%%%%%%%%%%%%%%%%%%%%%%
	\section{System Model}
	\label{sec:sm}
	%%%%%%%%%%%%%%%%%%%%%%%%%%%%%%%%%%%%%%%%%%%%%%%%%%%%%%%%%%%%%%%%%%%%%%%%%%%%%%%%
	In this section, we present the downlink MIMO system model.
	
	We consider an \emph{under-loaded or critically loaded} downlink MU-MIMO communication system with a base station (BS) employing $N$ transmit antennas, and $K$ users equipped with $M_k,k=1,\dots,K,$ antennas, such that
	\begin{align}
		N \geq \sum_{k=1}^{K} M_k.
	\end{align}
	The MIMO symbol vectors of the users are constructed using RS, as described in the following. First, the private message of the $k$-th user, $k=1,\dots,K,$ is encoded into a MIMO symbol vector $\boldsymbol{s}_{k} \in \mathbb{C}^{M_k\times 1}.$ We assume that $\mathrm{E}\mkern-\thinmuskip\left[\boldsymbol{s}_{k} \boldsymbol{s}_{k}^\mathrm{H}\right] = \boldsymbol{I}_{M_k}, \forall\,k.$ Additionally, a common message including messages to all downlink users is encoded into a MIMO symbol vector $\boldsymbol{s}_{\mathrm{c}} \in \mathbb{C}^{M\times 1}, \mathrm{E}\mkern-\thinmuskip\left[\boldsymbol{s}_{\mathrm{c}} \boldsymbol{s}_{\mathrm{c}}^\mathrm{H}\right] = \boldsymbol{I}_M,$ where $M = \mathrm{min}\mkern-\thinmuskip\left\{M_k,k=1,\dots,K\right\}.$ Symbol vectors $\boldsymbol{s}_{\mathrm{c}}$ and $\boldsymbol{s}_{k},\forall\,k,$ are assumed to be statistically independent.
	
	Next, the common MIMO symbol vector, $\boldsymbol{s}_{\mathrm{c}},$ and the private MIMO symbol vector, $\boldsymbol{s}_{k}, k=1,\dots,K,$ are precoded using linear precoders $\boldsymbol{P}_{\mathrm{c}} \in \mathbb{C}^{N\times M}$ and $\boldsymbol{P}_{k} \in \mathbb{C}^{N\times M_k},$ respectively, and the superimposed precoded symbol vectors are transmitted by the BS. The transmit power constraint at the BS is as follows:
	\begin{align}
		\mathrm{tr}\mkern-\thinmuskip\left(\boldsymbol{P}_{\mathrm{c}} \boldsymbol{P}_{\mathrm{c}}^\mathrm{H}\right) + \sum_{k = 1}^{K} \mathrm{tr}\mkern-\thinmuskip\left(\boldsymbol{P}_{k} \boldsymbol{P}_{k}^\mathrm{H}\right) \leq P_\mathrm{T}, \label{eqn:p1}
	\end{align}
	where $P_\mathrm{T}$ denotes the available transmit power.

	Let $\boldsymbol{H}_k \in \mathbb{C}^{M_k\times N}$ denote the MIMO channel matrix between the BS and user $k.$ Then, the received signal at user $k$ is given by
	\begin{align}
		\boldsymbol{y}_k = \boldsymbol{H}_k \Big(\boldsymbol{P}_{\mathrm{c}}\boldsymbol{s}_{\mathrm{c}} + \sum_{k' = 1}^{K} \boldsymbol{P}_{k'}\boldsymbol{s}_{k'}\Big) + \boldsymbol{z}_k,
	\end{align}
	where $\boldsymbol{z}_k \in \mathbb{C}^{M_k\times 1} \sim \mathcal{CN}(\boldsymbol{0}, \sigma^2\boldsymbol{I}_{M_k})$ denotes the complex additive white Gaussian noise (AWGN) vector at user $k.$ We assume that the users know their own MIMO channel matrices perfectly, and that the BS knows\footnote{We assume perfect channel knowledge at the BS and the users in order to obtain an upper bound on the achievable performance of the proposed SNS precoding and decoding schemes. The evaluation of the performance of the proposed design with imperfect channel knowledge is an interesting topic for future research.} all MIMO channel matrices $\boldsymbol{H}_k,k=1,\dots,K.$ Furthermore, we assume that the MIMO channel matrices have full row rank\footnote{We note that a row-rank deficient MIMO channel matrix can be transformed into a full row-rank matrix with fewer effective receive antennas via singular value decomposition, see e.g., \cite[App. C]{Scutari2009}.}.
	
	%%%%%%%%%%%%%%%%%%%%%%%%%%%%%%%%%%%%%%%%%%%%%%%%%%%%%%%%%%%%%%%%%%%%%%%%%%%%%%%%
	\section{Proposed Successive Null-Space Precoding and Decoding Schemes}
	\label{sec:pd}
	%%%%%%%%%%%%%%%%%%%%%%%%%%%%%%%%%%%%%%%%%%%%%%%%%%%%%%%%%%%%%%%%%%%%%%%%%%%%%%%%
	In this section, we present the proposed SNS precoding and decoding schemes. The proposed schemes utilize a fixed precoder structure based on the null spaces of the MIMO channels of the users.
	
	%%%%%%%%%%%%%%%%%%%%%%%%%%%%%%%%%%%%%%%%
	\subsection{Proposed Precoding Scheme}
	\label{sec:precoding}
	%%%%%%%%%%%%%%%%%%%%%%%%%%%%%%%%%%%%%%%%
	Let $\boldsymbol{N}_k \in \mathbb{C}^{N\times \left(N-\sum_{k' = 1}^{k-1} M_{k'}\right)}$ denote a matrix whose columns are the unit-length basis vectors of the null space of the following augmented matrix:
	\begin{align}
		\begin{bmatrix}
		\boldsymbol{H}_{1}^\mathrm{T} \bigm| \boldsymbol{H}_{2}^\mathrm{T} \bigm| \dots \bigm| \boldsymbol{H}_{k-1}^\mathrm{T}
		\end{bmatrix}^\mathrm{T}\negthickspace, \label{eqn:augh}
	\end{align}
	with the convention $\boldsymbol{N}_{1} = \boldsymbol{I}_N.$ The proposed \emph{SNS precoder} for user $k$ is constructed as follows:
	\begin{align}
		\boldsymbol{P}_{k} &= \boldsymbol{N}_{k} \boldsymbol{X}_{k}^\frac{1}{2}, \label{eqn:pk}
	\end{align}
	where 
	\begin{align}
		\boldsymbol{X}_{k} &\in \mathbb{C}^{\left(N-\sum_{k'=1}^{k-1}M_{k'}\right)\times \left(N-\sum_{k'=1}^{k-1}M_{k'}\right)}
	\end{align}
	is a symmetric, positive semi-definite matrix with rank $M_k,$ which is to be optimized. Since matrices $\boldsymbol{X}_{k},k=1,\dots,K,$ have rank $M_k,$ they can be factorized as:
	\begin{align}
		\boldsymbol{X}_{k} = \boldsymbol{X}_{k}^\frac{1}{2} (\boldsymbol{X}_{k}^\frac{1}{2})^\mathrm{H},\forall\,k,
	\end{align}
	where the rectangular matrix factors
	\begin{align}
		\boldsymbol{X}_{k}^\frac{1}{2} &\in \mathbb{C}^{\left(N-\sum_{k'=1}^{k-1}M_{k'}\right)\times M_k}
	\end{align}
	have $M_k$ columns each. Furthermore, we have
	\begin{align}
		\mathrm{tr}\mkern-\thinmuskip\left(\boldsymbol{P}_{k} \boldsymbol{P}_{k}^\mathrm{H}\right) &= \mathrm{tr}\mkern-\thinmuskip\left(\boldsymbol{N}_{k} \boldsymbol{X}_{k} \boldsymbol{N}_{k}^\mathrm{H}\right) \nonumber\\ 
		& = \text{tr}\Big({\underbrace{\boldsymbol{N}_{k}^\mathrm{H} \boldsymbol{N}_{k}}_{\boldsymbol{I}_{\left(N-\sum_{k'=1}^{k-1}M_{k'}\right)}} \boldsymbol{X}_{k}}\Big) 	= \mathrm{tr}\mkern-\thinmuskip\left(\boldsymbol{X}_{k}\right), \label{eqn:trsimp}
		\raisetag{1\normalbaselineskip}
	\end{align}
	for $k=1,\dots,K.$ Hence, the condition in (\ref{eqn:p1}) can be rewritten as:
	\begin{align}
		\mathrm{tr}\mkern-\thinmuskip\left(\boldsymbol{P}_{\mathrm{c}} \boldsymbol{P}_{\mathrm{c}}^\mathrm{H}\right) + \sum_{k = 1}^{K} \mathrm{tr}\mkern-\thinmuskip\left(\boldsymbol{X}_{k}\right) \leq P_\mathrm{T} \label{eqn:p2}.
	\end{align}
	
	\begin{remark}
		As $\boldsymbol{P}_{k}$ utilizes linear combinations of the basis vectors in $\boldsymbol{N}_{k},$ symbol vectors precoded with $\boldsymbol{P}_{k}$ do not cause IUI to users $k'=1,\dots,k-1,$ however, they cause IUI to the remaining users $k'=k+1,\dots,K.$ Furthermore, for user $k,$ the BD basis vectors \cite{Spencer2004} lie in a subspace spanned by the SNS basis vectors $\boldsymbol{N}_{k}.$ Hence, in $\boldsymbol{P}_{k},$ IUI can be adjusted by judiciously combining the basis vectors in $\boldsymbol{N}_{k}$ using $\boldsymbol{X}_{k}^\frac{1}{2}.$ In the extreme case, IUI can be completely eliminated by selecting only the BD subspace.
	\end{remark}
	
	\begin{remark}
		As seen from (\ref{eqn:augh}) and (\ref{eqn:pk}), the precoder matrices depend on the user labels $1,\dots,K.$ Users with lower indices have more degrees of freedom to choose their precoder $\boldsymbol{P}_{k}.$ Consequently, the user rates depend on the user labeling. Hence, user rate optimization should be carried out over all permutations of user labels.
	\end{remark}
	
	Based on the SNS precoder matrices given above, the received signal of user $k$ can be rewritten as follows:
	\begin{align}
		\boldsymbol{y}_k &= \boldsymbol{H}_k\boldsymbol{P}_{\mathrm{c}}\boldsymbol{s}_{\mathrm{c}} + \boldsymbol{H}_k \sum_{k'=1}^{k} \boldsymbol{N}_{k'} \boldsymbol{X}_{k'}^\frac{1}{2}\boldsymbol{s}_{k'} + \boldsymbol{z}_k. \label{eqn:y_r}
	\end{align}
	
	%%%%%%%%%%%%%%%%%%%%%%%%%%%%%%%%%%%%%%%%
	\subsection{Decoding Scheme}
	\label{sec:decoding}
	%%%%%%%%%%%%%%%%%%%%%%%%%%%%%%%%%%%%%%%%
	At user $k,$ decoding and SIC are performed as follows. First, using $\boldsymbol{y}_k,$ the common symbol vector $\boldsymbol{s}_{\mathrm{c}}$ is decoded treating the contributions of $\boldsymbol{s}_{k'},k'=1,\dots,k,$ as noise. Following successful decoding\footnote{Decoding is assumed to be always successful as the symbols are transmitted at or below their achievable rates. In practice, powerful codes that can closely approach these achievable rates can be utilized.}, the contribution of $\boldsymbol{s}_{\mathrm{c}}$ is eliminated, resulting in the signal:
	\begin{align}
		\boldsymbol{y}_k' &= \boldsymbol{y}_k - \boldsymbol{H}_k\boldsymbol{P}_{\mathrm{c}}\boldsymbol{s}_{\mathrm{c}} \nonumber\\&{}= \boldsymbol{H}_k \sum_{k'=1}^{k} \boldsymbol{N}_{k'} \boldsymbol{X}_{k'}^\frac{1}{2}\boldsymbol{s}_{k'} + \boldsymbol{z}_k. \label{eqn:y_r1}
	\end{align}
		
	Next, exploiting $\boldsymbol{y}_k',$ symbol vector $\boldsymbol{s}_{k}$ is decoded treating $\boldsymbol{s}_{k'},k'=1,\dots,k-1,$ as noise. The proposed SNS precoding and decoding strategies are shown schematically in Figure \ref{fig:decoding}.
	
	%%%%%%%%%%%%%%%%%%%%%%%%%%%%%%%%%%%%%%%%
	\subsection{Achievable Rate}
	%%%%%%%%%%%%%%%%%%%%%%%%%%%%%%%%%%%%%%%%
	Based on (\ref{eqn:y_r}) and (\ref{eqn:y_r1}), the achievable rate of user $k$ can be obtained as follows. The rate of MIMO symbol vector $\boldsymbol{s}_{k}$ is given by
	\begin{align}
		R_{k} = \log_2\det\Bigg(\boldsymbol{I}_{M_k} &+ \boldsymbol{H}_k\boldsymbol{Q}_{k}\boldsymbol{H}_k^\mathrm{H} \,\Big[\sigma^2\boldsymbol{I}_{M_k} \nonumber\\&{}+ \boldsymbol{H}_k \Big(\sum_{k'=1}^{k-1}\boldsymbol{Q}_{k'}\Big)\boldsymbol{H}_k^\mathrm{H}\Big]^{-1}\Bigg), \label{eqn:rk2}
	\end{align}
	where
	\begin{align}
		\boldsymbol{Q}_{k_1} &= \boldsymbol{N}_{k_1} \boldsymbol{X}_{k_1} \boldsymbol{N}_{k_1}^\mathrm{H},\label{eqn:qk}
	\end{align}
	for $k_1=1,\dots,K.$ Next, for the common MIMO symbol vector, the achievable rate at user $k$ is:
	\begin{align}
		R_{k,\mathrm{c}} = \log_2\det\Bigg(\boldsymbol{I}_{M_k} &+ \boldsymbol{H}_{k}\boldsymbol{Q}_{\mathrm{c}}\boldsymbol{H}_{k}^\mathrm{H} \,\Big[\sigma^2\boldsymbol{I}_{M_{k}} \nonumber\\&{}+ \boldsymbol{H}_{k} \Big(\sum_{k'=1}^{k}\boldsymbol{Q}_{k'}\Big)\boldsymbol{H}_{k}^\mathrm{H}\Big]^{-1}\Bigg),
	\end{align}
	where
	\begin{align}
		\boldsymbol{Q}_{\mathrm{c}} &= \boldsymbol{P}_{\mathrm{c}} \boldsymbol{P}_{\mathrm{c}}^\mathrm{H}. \label{eqn:qkc}
	\end{align}
	Since the common MIMO symbol vector must be decodeable at all users, its rate is chosen as:
	\begin{align}
		R_{\mathrm{c}} &= \mathrm{min}\mkern-\thinmuskip\left\{R_{k,\mathrm{c}}, k=1,\dots,K\right\}.
	\end{align}
	
	\begin{figure}
		\centering
		\begin{minipage}{0.5\textwidth}
			\centering
			\resizebox{\textwidth}{!}{%
				\renewcommand{\arraystretch}{1.5}
				\begin{tabular}{cc|c|c|c|ccc|c|cl}
					\hline
					Symbol/User      && $1$  & $2$ & $3$ &&        && $K$ && Precoder \\\hline\hline
					$\boldsymbol{s}_{\mathrm{c}}$ && \cellcolor{Turquoise}D    & \cellcolor{Turquoise}D   &  \cellcolor{Turquoise}D  &\cellcolor{Turquoise}&\cellcolor{Turquoise}$\cdots$&\cellcolor{Turquoise}& \cellcolor{Turquoise}D   && $\boldsymbol{P}_{\mathrm{c}}$ \\\hline
					$\boldsymbol{s}_{1}$ && \cellcolor{LimeGreen}S    & \cellcolor{Goldenrod}I   & \cellcolor{Goldenrod}I  &\cellcolor{Goldenrod}&\cellcolor{Goldenrod}$\cdots$&\cellcolor{Goldenrod}& \cellcolor{Goldenrod}I   && $\boldsymbol{X}_{1}^\frac{1}{2}$ \\\hline
					$\boldsymbol{s}_{2}$ &&      & \cellcolor{LimeGreen}S   &  \cellcolor{Goldenrod}I  &\cellcolor{Goldenrod}&\cellcolor{Goldenrod}$\cdots$&\cellcolor{Goldenrod}& \cellcolor{Goldenrod}I   && $\boldsymbol{N}_{2} \boldsymbol{X}_{2}^\frac{1}{2}$ \\\hline
					$\vdots$  &&      &     &     &&        &&     && $\vdots$ \\\hline
					$\boldsymbol{s}_{K}$ &&      &     &     &&&& \cellcolor{LimeGreen}S   && $\boldsymbol{N}_{K} \boldsymbol{X}_{K}^\frac{1}{2}$ \\\hline
			\end{tabular}}
			\medskip
			
			\resizebox{0.5\textwidth}{!}{%
				\begin{tabular}{|cl|}
					\hline
					\cellcolor{LimeGreen}S & Privately decoded symbol \\
					\cellcolor{Turquoise}D & Commonly decoded symbol \\
					\cellcolor{Goldenrod}I & Symbols treated as noise\\
					\hline
			\end{tabular}}
			\caption{Schematic diagram of the proposed SNS precoding and decoding schemes.}
			\label{fig:decoding}
		\end{minipage}
	\end{figure}
	
	%%%%%%%%%%%%%%%%%%%%%%%%%%%%%%%%%%%%%%%%
	\section{WSR Maximization}
	\label{sec:mwsr}
	%%%%%%%%%%%%%%%%%%%%%%%%%%%%%%%%%%%%%%%%
	In this section, we formulate the WSR optimization problem. Since the problem is highly non-convex, we characterize the maximum WSR through a feasible lower bound obtained via SCA.
	
	\subsection{Weighted Sum Rate}
	Let $0 \leq \eta_k \leq 1,k=1,\dots,K,$ $\sum_{k=1}^{K}\eta_k = 1,$ denote fixed weights which can be chosen to adjust the rates of the users during power allocation \cite[Sec. 4]{WangGiannakis2011}. In our case, these fixed weights can also be utilized to adjust the fractional rates of the users for the common message by assigning a fraction $\eta_k$ of the available bits in the common message to user $k.$ Then, the WSR is given by
	\begin{align}
		R_\mathrm{wsr} = \sum_{k=1}^{K} \eta_k^2 R_{\mathrm{c}} + \sum_{k = 1}^{K} \eta_k R_{k}. \label{eqn:wsr}
	\end{align}
	
	\subsection{Problem Formulation}
	Based on (\ref{eqn:wsr}), the maximum WSR is the solution $R_\mathrm{wsr}^\star$ of the following optimization problem:
	\begin{maxi!}
		{\substack{\boldsymbol{Q}_{\mathrm{c}} \succcurlyeq \boldsymbol{0},\\\boldsymbol{X}_{k} \succcurlyeq \boldsymbol{0},\forall\,k}}
		{R_\mathrm{wsr}}
		{\label{opt:wsr}}
		{\mathclap{R_\mathrm{wsr}^\star =}\nonumber\\}
		\addConstraint{\text{C1: } \mathrm{tr}\mkern-\thinmuskip\left(\boldsymbol{Q}_{\mathrm{c}}\right) + \sum_{k = 1}^{K} \mathrm{tr}\mkern-\thinmuskip\left(\boldsymbol{X}_{k}\right)\leq P_{\mathrm{T}}\label{cons:t1}}{}{}
		\addConstraint{\text{C2: } \mathrm{rank}\mkern-\thinmuskip\left(\boldsymbol{X}_{k}\right)\leq M_k,\quad\forall\,k\label{cons:rank2}}{}{}
		\addConstraint{\text{C3: } \mathrm{rank}\mkern-\thinmuskip\left(\boldsymbol{Q}_{\mathrm{c}}\right)\leq M,\label{cons:rankc}}{}{}
	\end{maxi!}
	where the WSR is maximized for all permutations of user labels. Solving (\ref{opt:wsr}) entails a very high computational complexity due to the non-convex objective function and the rank constraints in (\ref{cons:rank2}) and (\ref{cons:rankc}). Hence, in the following, we develop a tractable suboptimal solution for (\ref{opt:wsr}), which also provides a feasible lower bound for $R_\mathrm{wsr}^\star.$
	
	\subsection{Suboptimal Solution}
	As a first step, we relax (\ref{opt:wsr}) by eliminating the rank constraints (\ref{cons:rank2}) and (\ref{cons:rankc}) to obtain a relaxed optimization problem, which we denote by (\ref{opt:wsr}-). However,  (\ref{opt:wsr}-) is still non-convex due to the non-convex objective function $R_\mathrm{wsr}.$ Nevertheless, a locally optimal solution for (\ref{opt:wsr}-) can now be obtained via SCA \cite{Razaviyayn2014}. The SCA procedure for (\ref{opt:wsr}-) is described in detail in the following section. Yet, as a solution for (\ref{opt:wsr}-) may not satisfy rank constraints (\ref{cons:rank2}) and (\ref{cons:rankc}), it may not be a feasible solution for (\ref{opt:wsr}).
	
	Hence, next, \emph{based on the obtained locally optimal solution,} we reformulate (\ref{opt:wsr}-) with new optimization variables which automatically satisfy the rank constraints in (\ref{cons:rank2}) and (\ref{cons:rankc}). The resulting solution for the new optimization problem is a feasible solution for (\ref{opt:wsr}). The reformulated problem is presented in Section \ref{sec:ref}.
	
	\subsection{Successive Convex Approximation}
	\label{sec:sca}
	For solving (\ref{opt:wsr}-) via SCA, an inner convex optimization problem based on a first-order approximation of the non-convex objective function $R_{\mathrm{wsr}}$ is constructed. This inner optimization problem is solved repeatedly until convergence, upto a numerical tolerance $\epsilon.$ In each iteration, based on the obtained optimal solution, the first-order approximation is updated and used as the objective function for the next iteration. The procedure is described in detail in the following.
	
	In iteration $l=1,2,\dots,$ a convex approximation of the objective function $R_{\mathrm{wsr}},$ denoted by $\tilde{R}_{\mathrm{wsr}},$ is constructed based on a first-order approximation around \emph{given points} $\breve{\boldsymbol{X}}_{k}^{(l-1)} \in \mathbb{C}^{\left(N-\sum_{k'=1}^{k-1}M_{k'}\right)\times \left(N-\sum_{k'=1}^{k-1}M_{k'}\right)}, k=1,\dots,K,$ with initial values $\breve{\boldsymbol{X}}_{k}^{(0)} = \boldsymbol{0}, k=1,\dots,K,$ as follows:
	\begin{align}
		\tilde{R}_\mathrm{wsr} = \sum_{k=1}^{K} \eta_k^2 \tilde{R}_{\mathrm{c}} + \sum_{k = 1}^{K} \eta_k \tilde{R}_{k},
	\end{align}
	where $\tilde{R}_{\mathrm{c}} = \mathrm{min}\mkern-\thinmuskip\left\{\tilde{R}_{k,\mathrm{c}}, k=1,\dots,K\right\},$ and $\tilde{R}_{k,\mathrm{c}}$ and $\tilde{R}_{k},k=1,\dots,K,$ are given in (\ref{eqn:trc}) and (\ref{eqn:tr12}), on top of the next page, respectively.
	\begin{figure*}
		\begin{align}
			\tilde{R}_{k,\mathrm{c}} = {}&\mathrm{log}_2\mathrm{\,det}\mkern-\thinmuskip\left(\boldsymbol{I}_{M_k} + \frac{1}{\sigma^2}\boldsymbol{H}_{k}\boldsymbol{Q}_{\mathrm{c}}\boldsymbol{H}_{k}^\mathrm{H} + \frac{1}{\sigma^2}\boldsymbol{H}_{k} \Big(\sum_{k'=1}^{k}\boldsymbol{Q}_{k'}\Big)\boldsymbol{H}_{k}^\mathrm{H}\right) -\mathrm{log}_2\mathrm{\,det}\mkern-\thinmuskip\left(\boldsymbol{I}_{M_k} + \frac{1}{\sigma^2}\boldsymbol{H}_{k} \breve{\boldsymbol{Q}}_{k+1}^{(l-1)} \boldsymbol{H}_{k}^\mathrm{H}\right) \nonumber\\&{} - \frac{1}{\sigma^2\log_{\mathrm{e}}(2)}\sum_{k'=1}^{k-1}\mathrm{tr}\mkern-\thinmuskip\left(\boldsymbol{N}_{k'}^\mathrm{H} \boldsymbol{H}_k^\mathrm{H} \Big[\boldsymbol{I}_{M_k} + \frac{1}{\sigma^2}\boldsymbol{H}_k \breve{\boldsymbol{Q}}_{k+1}^{(l-1)} \boldsymbol{H}_k^\mathrm{H}\Big]^{-1} \boldsymbol{H}_k \boldsymbol{N}_{k'}\Big(\boldsymbol{X}_{k'}-\breve{\boldsymbol{X}}_{k'}^{(l-1)}\Big)\right), \label{eqn:trc} \\
			\tilde{R}_{k} = {}&\mathrm{log}_2\mathrm{\,det}\mkern-\thinmuskip\left(\boldsymbol{I}_{M_k} + \frac{1}{\sigma^2}\boldsymbol{H}_k\boldsymbol{Q}_{k}\boldsymbol{H}_k^\mathrm{H} + \frac{1}{\sigma^2}\boldsymbol{H}_k \Big(\sum_{k'=1}^{k-1}\boldsymbol{Q}_{k'}\Big)\boldsymbol{H}_k^\mathrm{H}\right) - \mathrm{log}_2\mathrm{\,det}\mkern-\thinmuskip\left(\boldsymbol{I}_{M_k} + \frac{1}{\sigma^2}\boldsymbol{H}_k \breve{\boldsymbol{Q}}_k^{(l-1)} \boldsymbol{H}_k^\mathrm{H}\right) \nonumber\\&{}- \frac{1}{\sigma^2\log_{\mathrm{e}}(2)}\sum_{k'=1}^{k-1}\mathrm{tr}\mkern-\thinmuskip\left(\boldsymbol{N}_{k'}^\mathrm{H} \boldsymbol{H}_k^\mathrm{H} \Big[\boldsymbol{I}_{M_k} + \frac{1}{\sigma^2}\boldsymbol{H}_k \breve{\boldsymbol{Q}}_k^{(l-1)} \boldsymbol{H}_k^\mathrm{H}\Big]^{-1}\boldsymbol{H}_k \boldsymbol{N}_{k'}\Big(\boldsymbol{X}_{k'}-\breve{\boldsymbol{X}}_{k'}^{(l-1)}\Big)\right) \label{eqn:tr12}
		\end{align}
		\hrulefill
		\vspace{-0.3cm}
	\end{figure*}%
	In (\ref{eqn:trc}) and (\ref{eqn:tr12}), we used:
	\begin{align}
		\breve{\boldsymbol{Q}}_{k_1}^{(l-1)} &= \sum_{k'=1}^{k_1-1}\boldsymbol{N}_{k'} \breve{\boldsymbol{X}}_{k'}^{(l-1)} \boldsymbol{N}_{k'}^\mathrm{H},
	\end{align}
	for $k_1 = 1,\dots,K.$ Moreover, the first-order approximations given in (\ref{eqn:trc}) and (\ref{eqn:tr12}) are obtained based on matrix derivatives, see \cite{Petersen2012}. Next, an inner convex optimization problem with $\tilde{R}_{\mathrm{wsr}}$ as the objective function is constructed as follows:
	\begin{maxi!}
		{\substack{\boldsymbol{Q}_{\mathrm{c}} \succcurlyeq \boldsymbol{0},\\\boldsymbol{X}_{k} \succcurlyeq \boldsymbol{0},\forall\,k}}
		{\tilde{R}_\mathrm{wsr}}
		{\label{opt:wsrlo}}
		{\tilde{R}_\mathrm{wsr}^\star =}
		\addConstraint{{\text{C1}}\text{.}}{}{}
	\end{maxi!}
	
	The inner convex optimization problem (\ref{opt:wsrlo}) is solved using standard convex optimization tools \cite{Boyd2004} to obtain the optimal value, $\tilde{R}_\mathrm{wsr}^{\star(l)},$ and the corresponding optimal solution $\boldsymbol{Q}_{\mathrm{c}}^\star,\boldsymbol{X}_{k}^\star,\forall\,k.$ The obtained solution is used as the given point for the next iteration, i.e., $\breve{\boldsymbol{X}}_{k}^{(l)} = \boldsymbol{X}_{k}^\star.$ This process gradually tightens the first-order approximation of the objective function around a local optimum of (\ref{opt:wsr}-). Hence, the corresponding sequence of the optimal values of (\ref{opt:wsrlo}), $\tilde{R}_\mathrm{wsr}^{\star(l)},l=1,2,\dots,$ converges to a local optimum of (\ref{opt:wsr}-) \cite{Razaviyayn2014}. In our case, the iterations are continued until convergence upto a numerical tolerance $\epsilon.$ The algorithm is summarized in Algorithm \ref{alg:wsrlo}. Furthermore, similarly to (\ref{opt:wsr}-), Algorithm  \ref{alg:wsrlo} is applied for all permutations of user labels, and the maximal  $\tilde{R}_\mathrm{wsr}^\star$ and the corresponding solution are chosen for reformulating (\ref{opt:wsr}-), as described below.
	
	\begin{figure}
		\vspace{-0.25cm}
		\begin{algorithm}[H]
			\small
			\begin{algorithmic}[1]
				\STATE {Initialize $\breve{\boldsymbol{X}}_{k}^{(0)} = \boldsymbol{0},\forall\,k,$ numerical tolerance $\epsilon,$ and iteration index $l=0.$}
				\REPEAT	
				\STATE {$l \leftarrow l + 1$}
				\STATE {Update the first-order approximations in (\ref{eqn:trc}) and (\ref{eqn:tr12}) based on $\breve{\boldsymbol{X}}_{k}^{(l-1)}.$}
				\STATE {Solve convex optimization problem (\ref{opt:wsrlo}) to obtain the optimal value $\tilde{R}_\mathrm{wsr}^{\star(l)}$ and solution $\boldsymbol{Q}_{\mathrm{c}}^\star, \boldsymbol{X}_{k}^\star,\forall\,k.$}
				\STATE {Set $\breve{\boldsymbol{X}}_{k}^{(l)} = \boldsymbol{X}_{k}^\star,\forall\,k.$}
				\UNTIL {$l > 1$ and $|\tilde{R}_\mathrm{wsr}^{\star(l)} - \tilde{R}_\mathrm{wsr}^{\star(l-1)}| < \epsilon$}
				\STATE{Return $\tilde{R}_\mathrm{wsr}^\star = \tilde{R}_\mathrm{wsr}^{\star(l)}$ and $\boldsymbol{Q}_{\mathrm{c}}^\star, \boldsymbol{X}_{k}^\star,\forall\,k,$ as the optimal value and the corresponding solution.}
			\end{algorithmic}
			\caption{Algorithm for solving (\ref{opt:wsr}-) via SCA.}
			\label{alg:wsrlo}
		\end{algorithm}
		\vspace{-0.5cm}
	\end{figure}
	
	\subsection{Problem Reformulation}
	\label{sec:ref}
	In the following, we reformulate (\ref{opt:wsr}-) with new optimization variables. Let $\boldsymbol{Q}_{\mathrm{c}}^\star,$ $\boldsymbol{X}_{k}^\star,k=1,\dots,K,$ denote the locally optimal solution of (\ref{opt:wsr}-) obtained with Algorithm \ref{alg:wsrlo}. We use this solution to define new matrix structures for $\boldsymbol{Q}_{\mathrm{c}}$ and $\boldsymbol{Q}_{k},k=1,\dots,K,$ to replace the ones in (\ref{eqn:qk}) and (\ref{eqn:qkc}), as follows:
	\begin{align}
		\boldsymbol{Q}_{\mathrm{c}} &= \boldsymbol{U}_{\mathrm{c}} \tilde{\boldsymbol{X}}_{\mathrm{c}} \boldsymbol{U}_{\mathrm{c}}^\mathrm{H}, \label{eqn:qkc'} \\
		\boldsymbol{Q}_{k} &= \boldsymbol{N}_{k} \boldsymbol{U}_{k} \tilde{\boldsymbol{X}}_{k} \boldsymbol{U}_{k}^\mathrm{H} \boldsymbol{N}_{k}^\mathrm{H}, \label{eqn:qk'}
	\end{align}
	where $\boldsymbol{U}_{\mathrm{c}} \in \mathbb{C}^{N\times M}$ and $\boldsymbol{U}_k \in \mathbb{C}^{\left(N-\sum_{k'=1}^{k-1}M_{k'}\right)\times M_k}$ contain the eigenvectors of $\boldsymbol{Q}_{\mathrm{c}}^\star$ and $\boldsymbol{X}_{k}^\star,$ respectively, corresponding to their $M$ and $M_k$ largest eigenvalues, respectively. We follow the convention that the eigenvectors of a zero matrix are zero vectors. Here, $\tilde{\boldsymbol{X}}_{\mathrm{c}} \in \mathbb{C}^{M\times M}$ and $\tilde{\boldsymbol{X}}_{k} \in \mathbb{C}^{M_k\times M_k},\forall\,k,$ are symmetric, positive semi-definite matrices which form the new optimization variables.
	
	\begin{remark}
		The rank of $\boldsymbol{Q}_{\mathrm{c}}$ in (\ref{eqn:qkc'}) is limited to $M$ as $\boldsymbol{Q}_{\mathrm{c}}$ is the product of $N\times M,$ $M\times M,$ and $M\times N$ matrices $\boldsymbol{U}_{\mathrm{c}},$ $\tilde{\boldsymbol{X}}_{\mathrm{c}},$ and $\boldsymbol{U}_{\mathrm{c}}^\mathrm{H},$ respectively. Similarly, the rank of $\boldsymbol{Q}_{k}$ in (\ref{eqn:qk'}) is limited to $M_k.$
	\end{remark}
		
	Problem (\ref{opt:wsr}-) can now be reformulated in terms of the new matrix structures as follows:
	\begin{maxi!}
		{\substack{\tilde{\boldsymbol{X}}_{\mathrm{c}} \succcurlyeq \boldsymbol{0},\\\tilde{\boldsymbol{X}}_{k} \succcurlyeq \boldsymbol{0},\forall\,k}}
		{R_\mathrm{wsr}}
		{\label{opt:wsrlb}}
		{\mathclap{R_\mathrm{wsrlb}^\star =}\nonumber\\}
		\addConstraint{\overline{\text{C1}}\text{: }\mathrm{tr}\mkern-\thinmuskip\left(\tilde{\boldsymbol{X}}_{\mathrm{c}}\right) + \sum_{k = 1}^{K} \mathrm{tr}\mkern-\thinmuskip\left(\tilde{\boldsymbol{X}}_{k}\right) \leq P_{\mathrm{T}},\label{cons:t1lb}}{}{}		
	\end{maxi!}
	where the power constraint in (\ref{cons:t1lb}) is obtained analogously to (\ref{eqn:trsimp}), and as earlier, the WSR is maximized for all permutations of user labels. A locally optimal solution for (\ref{opt:wsrlb}) can be obtained via SCA analogously to the procedure in Section \ref{sec:sca} for (\ref{opt:wsr}-), i.e., Algorithm \ref{alg:wsrlo} is applicable where the optimization variables of (\ref{opt:wsrlo}) and the first-order approximations in (\ref{eqn:trc}) and (\ref{eqn:tr12}) are revised to account for the new variables and matrix structures in (\ref{eqn:qkc'}) and (\ref{eqn:qk'}).
	
	\begin{remark}
		A locally optimal solution of (\ref{opt:wsrlb}) is a feasible lower bound for $R_\mathrm{wsr}^\star$ in (\ref{opt:wsr}). The solution is feasible because the matrix structures in (\ref{eqn:qkc'}) and (\ref{eqn:qk'}) automatically ensure rank constraints (\ref{cons:rank2}) and (\ref{cons:rankc}). However, the solution is suboptimal because the optimization variables in (\ref{opt:wsrlb}) have fewer degrees of freedom compared to those in (\ref{opt:wsr}).
	\end{remark}	
			
	%%%%%%%%%%%%%%%%%%%%%%%%%%%%%%%%%%%%%%%%%%%%%%%%%%%%%%%%%%%%%%%%%%%%%%%%%%%%%%%%
	\section{Simulation Results}
	\label{sec:sim}	
	%%%%%%%%%%%%%%%%%%%%%%%%%%%%%%%%%%%%%%%%%%%%%%%%%%%%%%%%%%%%%%%%%%%%%%%%%%%%%%%%
	
	\begin{figure}
		\centering
		\begin{minipage}[t]{0.48\textwidth}
			\centering
			\includegraphics[width=0.9\textwidth,keepaspectratio]{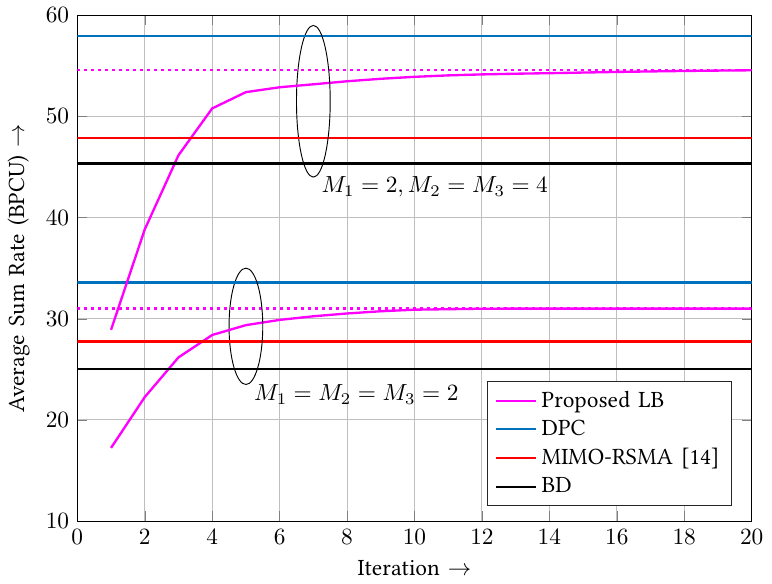}
			\caption{Convergence of Algorithm \ref{alg:wsrlo} for $P_\mathrm{T} = 20 \text{ dBm.}$}
			\label{fig:c244_1}
		\end{minipage}
		\vspace{-0.3cm}
	\end{figure}
	
	\begin{figure*}
		\centering
		\begin{minipage}[t]{0.48\textwidth}
			\centering
			\includegraphics[width=0.9\textwidth,keepaspectratio]{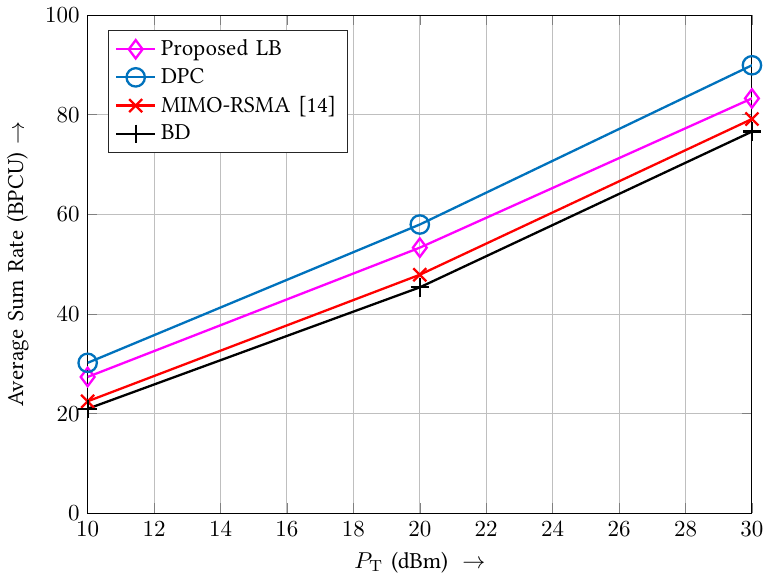}
			\caption{Average WSR for $M_1=2,M_2=M_3=4,N=10,$ and $d_1,d_2,d_3 = 50 \text{ m.}$}
			\label{fig:244_1}
		\end{minipage}%
  	    \hspace{0.02\textwidth}%
		\begin{minipage}[t]{0.48\textwidth}
			\centering
			\includegraphics[width=0.9\textwidth,keepaspectratio]{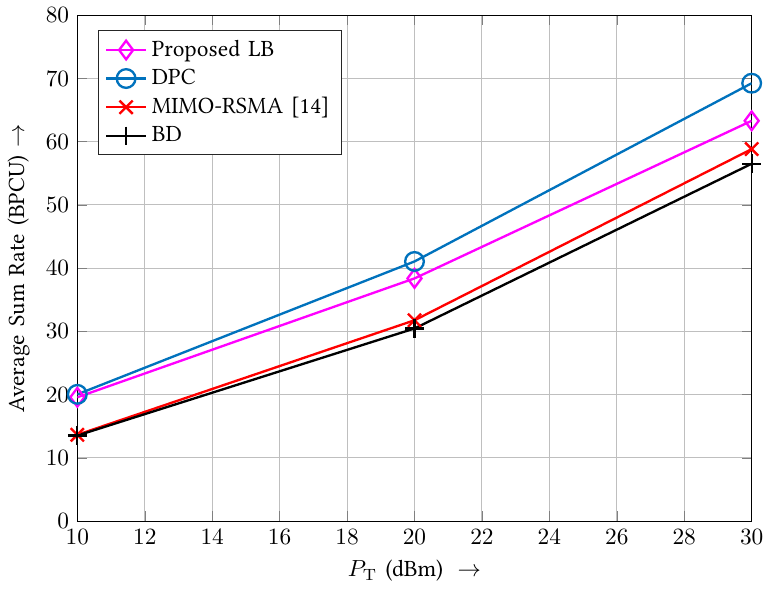}
			\caption{Average WSR for $M_1=2,M_2=M_3=4,N=10,$ and $d_1,d_2,d_3 = 250,150,50 \text{ m.}$}
			\label{fig:244_2}
		\end{minipage}
	\end{figure*}
	
	In this section, we evaluate the performance of the proposed SNS precoding and decoding schemes via simulations. For our simulations, we consider the critically loaded case, i.e., $N=\sum_{k = 1}^{K} M_k,$ and set the noise variance to $\sigma^2 = -35 \text{ dBm.}$ Furthermore, the channels $\boldsymbol{H}_k,k=1,\dots,K,$ are modeled as:
	\begin{align}
		\boldsymbol{H}_k = \frac{1}{\sqrt{L_k}} \boldsymbol{G}_k,
	\end{align}
	where the elements of the matrix $\boldsymbol{G}_k \sim \mathbb{C}^{M_k\times N}, k=1,\dots,K,$ are drawn from independent and identically distributed random variables $[\boldsymbol{G}_k]_{ij} \sim \mathcal{CN}(0,1),\forall\,i,j,k.$ The scalar $L_k$ models the path loss of user $k,$ and is set to $d_k^2,$ where $d_k$ denotes the distance (in meters) of user $k$ from the BS.	
	
	In the presented figures, we consider a system with three users, i.e., $K=3.$ The user weights are set to be equal, i.e., $\eta_k = 1/K,k=1,\dots,K,$ in order to obtain the maximum sum rate. The precoding vectors and power allocation for the lower bound on the WSR of the proposed design are obtained as described in Section \ref{sec:mwsr}. These are then used to compute the sum rate. The sum rate is averaged over multiple MIMO channel realizations so as to obtain a $99\%$ confidence interval of $\pm 1$ bit per channel use (BPCU) for the average sum rate. The obtained average sum rate is compared with the average sum rates of DPC, obtained by exploiting the BC-MAC duality \cite{Vishwanath2003}, MIMO-RSMA \cite{Flores2019}, and BD \cite{Spencer2004}. Furthermore, in Algorithm \ref{alg:wsrlo}, the numerical tolerance is set to $\epsilon = 10^{-5}.$
	
	First, in Figure \ref{fig:c244_1}, we study the convergence of Algorithm \ref{alg:wsrlo} for antenna configurations $M_1=M_2=M_3=2, N=6,$ and $M_1=2, M_2=M_3=4, N=10.$ The transmit power budget is set to $P_\mathrm{T} = 20 \text{ dBm.}$ The user distances from the BS are set to be equal, i.e., $d_k=50 \text{ m},k=1,2,3.$ From the figure, we observe that, for the considered cases, the average sum rate converges in about $20$ iterations. Furthermore, comparing the convergence speed for the two antenna configurations, we note that the convergence speed decreases only moderately with an increase in the number of variables.
	
	Next, in Figures \ref{fig:244_1} and \ref{fig:244_2}, shown on top of the next page, we study the obtained average sum rate for $M_1=2, M_2=M_3=4, N=10.$ In Figure \ref{fig:244_1}, the user distances from the BS are set to be equal, i.e., $d_k=50 \text{ m},k=1,2,3,$ and in Figure \ref{fig:244_2}, the user distances are unequal, i.e., $d_1,d_2,d_3 = 250, 150, 50 \text{ m}.$ From the figures, we observe that the proposed SNS scheme outperforms both MIMO-RSMA \cite{Flores2019} and BD as the proposed SNS precoder design can adjust the IUI to enhance the sum rate. At low-to-medium SNRs and when the user distances are dissimilar, the proposed SNS scheme has a relatively small gap to DPC. However, at high SNRs, the gap to DPC is larger. This is because, at high SNRs, the system is interference limited and the most beneficial strategy is to decode all substantial interference.

	%%%%%%%%%%%%%%%%%%%%%%%%%%%%%%%%%%%%%%%%%%%%%%%%%%%%%%%%%%%%%%%%%%%%%%%%%%%%%%%%
	\section{Conclusion}
	\label{sec:con}
	%%%%%%%%%%%%%%%%%%%%%%%%%%%%%%%%%%%%%%%%%%%%%%%%%%%%%%%%%%%%%%%%%%%%%%%%%%%%%%%%
	In this paper, we considered the precoder design for an under-loaded or critically loaded downlink MU-MIMO communication system employing RS at the transmitter and single-stage SIC at the receivers. We proposed the SNS precoding and decoding schemes which utilize linear combinations of the null-space basis vectors of the successively augmented MIMO channel matrices of the users as precoding vectors to adjust the IUI. Furthermore, we formulated the WSR maximization problem and obtained a feasible lower bound for the maximum WSR of the proposed SNS scheme via SCA. Our simulation results reveal that the obtained lower bound outperforms MIMO-RSMA \cite{Flores2019} and BD, and has a relatively small gap to DPC, especially at low to medium SNRs and when the user distances are dissimilar.
	%%%%%%%%%%%%%%%%%%%%%%%%%%%%%%%%%%%%%%%%%%%%%%%%%%%%%%%%%%%%%%%%%%%%%%%%%%%%%%%%
	% Appendices
	%%%%%%%%%%%%%%%%%%%%%%%%%%%%%%%%%%%%%%%%%%%%%%%%%%%%%%%%%%%%%%%%%%%%%%%%%%%%%%%%
	% \begin{appendices}
	% Grab forced line break - \\* - and replace with :
	\renewcommand{\thesection}{\Alph{section}}
	\renewcommand{\thesubsection}{\thesection.\arabic{subsection}}
	\renewcommand{\thesectiondis}[2]{\Alph{section}:}
	\renewcommand{\thesubsectiondis}{\thesection.\arabic{subsection}:}
	%%%%%%%%%%%%%%%%%%%%%%%%%%%%%%%%%%%%%%%%%%%%%%%%%%%%%%%%%%%%%%%%%%%%%%%%%%%%%%%%
	%%%%%%%%%%%%%%%%%%%%%%%%%%%%%%%%%%%%%%%%%%%%%%%%%%%%%%%%%%%%%%%%%%%%%%%%%%%%%%%%
	% \end{appendices}
	
	%%%%%%%%%%%%%%%%%%%%%%%%%%%%%%%%%%%%%%%%%%%%%%%%%%%%%%%%%%%%%%%%%%%%%%%%%%%%%%%%
	% Bibliography
	%%%%%%%%%%%%%%%%%%%%%%%%%%%%%%%%%%%%%%%%%%%%%%%%%%%%%%%%%%%%%%%%%%%%%%%%%%%%%%%%
	\bibliographystyle{IEEEtran}
	\bibliography{references}
	
\end{document}